\documentclass[sigconf, nonacm]{acmart}

\usepackage{amsmath,amsfonts}

\usepackage{amssymb}
\usepackage{stfloats}
\usepackage{algorithmic}
\usepackage{graphicx}
\usepackage{textcomp}
\usepackage{xcolor}
\usepackage{booktabs}
\usepackage{multirow}
\usepackage{enumitem}
\usepackage{subcaption}
\usepackage{hyperref}
\usepackage{url}
\usepackage{microtype}
\usepackage{xspace}

\usepackage{booktabs}
\usepackage{array}
\usepackage{multirow}
\usepackage{xcolor}
\usepackage{colortbl}
\usepackage{fontawesome5}
\usepackage{graphicx}

\definecolor{RampBlue}{RGB}{33,84,166}
\definecolor{RampDark}{RGB}{30,34,42}
\definecolor{RampRed}{RGB}{210,55,55}
\definecolor{RampGreen}{RGB}{35,135,85}
\definecolor{RampOrange}{RGB}{225,145,35}
\definecolor{RampGray}{RGB}{248,249,252}
\definecolor{RampLightBlue}{RGB}{236,242,252}

\definecolor{darkblue}{rgb}{0, 0, 0.5}
\hypersetup{colorlinks=true, citecolor=darkblue, linkcolor=darkblue, urlcolor=darkblue}

\newcommand{\ramp}{\textsc{Ramp}\xspace}
\newcommand{\yatcc}{\textit{YatCC}\xspace}

\newcommand{\resurrection}{\textit{resurrection}\xspace}
\newcommand{\cnt}{15\xspace}

\newcommand{\myt}[1]{\textit{T{#1}}\xspace}
\newcommand{\myT}[1]{\textit{Task{#1}}\xspace}
\newcommand{\aei}[1]{\textit{AEI}\xspace}

\newcommand{\xwz}[2]{\textcolor{black}{#1}} 

\newif\ifreview
\reviewtrue

\ifreview
\fi

\reviewtrue

\title{Benchmarks are Not Enough: \ramp for \underline{R}untime \underline{A}ssessing of Agentic \underline{M}odels in \underline{P}roduction Systems$^{*}$}

\AtBeginDocument{%
  }

\setcopyright{acmlicensed}
\copyrightyear{2018}
\acmYear{2018}
\acmDOI{XXXXXXX.XXXXXXX}
\acmConference[Conference acronym 'XX]{Make sure to enter the correct
  conference title from your rights confirmation email}{June 03--05,
  2018}{Woodstock, NY}
\acmISBN{978-1-4503-XXXX-X/2018/06}

\definecolor{deepgreen}{HTML}{15803D}
\definecolor{deepred}{HTML}{B91C1C}
\definecolor{deepblue}{HTML}{1D4ED8}

\begin{document}

\begin{teaserfigure}
\centering
\includegraphics[width=\textwidth]{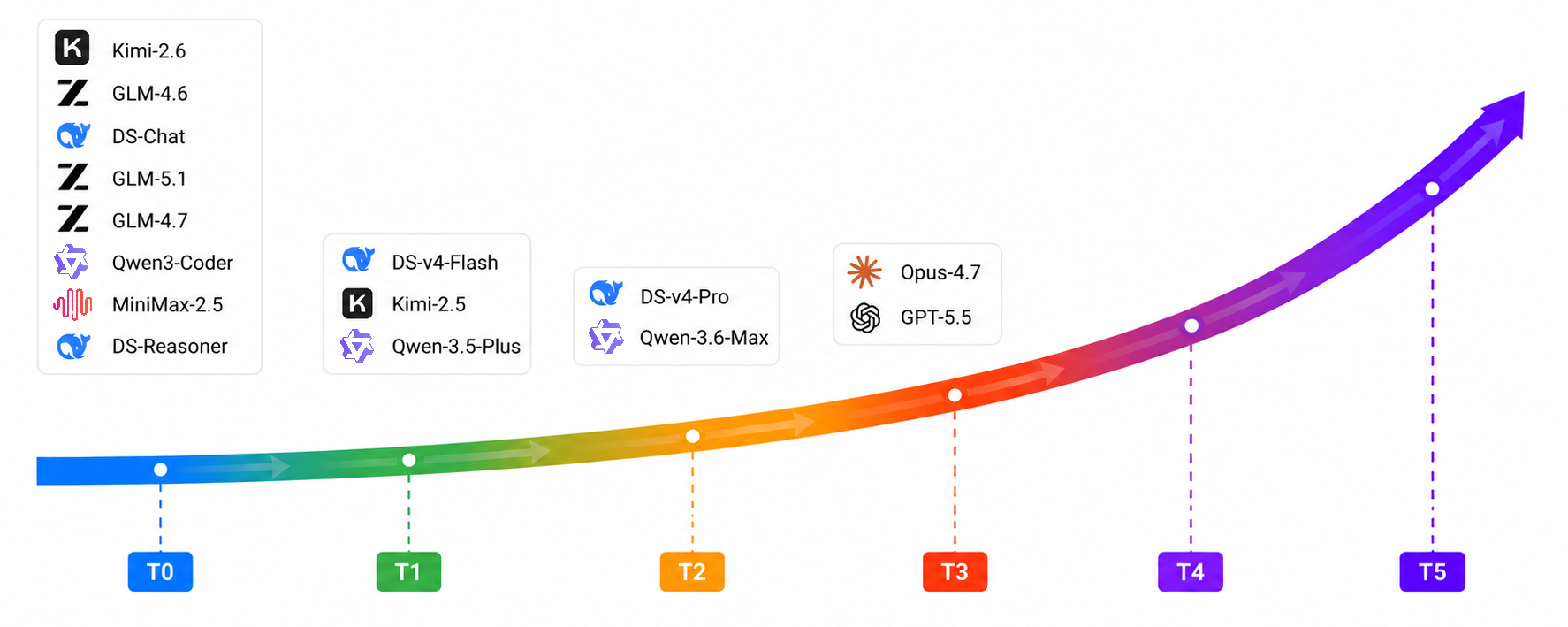}
\caption*{How Far Can Today’s Agentic Models Go on the \ramp to Real System Development?
\myt{0}–\myt{5} correspond to increasingly sophisticated implementation tasks from the \yatcc{} production infrastructure project. A model reaches tier \myt{x} only if it correctly and completely accomplishes every task up to \myt{x}, exposing the practical limits of current LLMs in sustained multi-stage software development.
}
\label{fig:llm-tier}
\vspace{0.25in}
\end{teaserfigure}

\author{Yipeng Ouyang,\hspace{0.4em} Xin Huang,\hspace{0.4em} Bingjie Liu,\hspace{0.4em} Zhongchun Zheng,\hspace{0.4em} Yuhao Gu,\hspace{0.4em} Xianwei Zhang$^{\dagger}$}

\affiliation{%
  \institution{School of Computer Science and Engineering, Sun Yat-sen University, Guangzhou, China \hspace{0.5em}}
  \country{}
}

\affiliation{%
  \institution{\{ouyyp5, huangx385, liubj8, zhengzhch3, guyh9\}@mail2.sysu.edu.cn, \{zhangxw79\}@mail.sysu.edu.cn}
  \country{}
}

\affiliation{%
    \institution{\url{http://ramp.yatcc-ai.com/}}
    \country{}
}

\thanks{$^{*}$ Timeline: Idea proposed on April 11th, 2026; Project initiated on April 22nd; Repository and leaderboard open-sourced on May 10th; Draft completed on May 14th.}
\thanks{$^{\dagger}$ Corresponding author.}

\begin{abstract}
Large language model (LLM) agents are rapidly evolving from coding assistants into autonomous software engineering systems. \xwz{However, existing}{yet} evaluation methodologies remain \xwz{largely centered on}{rooted in} static, isolated, \xwz{and}{} short-horizon benchmarks that fail to capture the dynamic complexity of real-world production workflows. \xwz{As a result, benchmark performance may poorly reflect practical capability under realistic runtime environments invovling long execution chains, tool interactions, dependency management, and iterative feedback loops.}{}
We thus present \ramp ({\it Runtime Assessment of Models in Production}), a production-grounded infrastructure for assessing long-horizon software engineering agents. Built upon the \yatcc{} integrated platform, \ramp provides a unified runtime assessment architecture supporting heterogeneous LLM providers and agent SDKs through standardized orchestration and execution interfaces. \ramp introduces realistic compiler-construction workloads with serial dependencies and complex toolchain interactions, together with a staged recovery mechanism for analyzing execution behavior under partial workflow failure. The framework further incorporates utility-oriented multi-dimensional metrics that jointly evaluate outcome quality and process efficiency.
We conduct runtime assessments across \cnt{} mainstream models and observe substantial capability degradation that remains largely invisible to conventional isolated benchmarks.
Task completion rates progressively collapse across serial workflows, dropping from 100\% in the initial stage to only 20.0\% in the final stage, while none of the evaluated models successfully completes the entire pipeline. Runtime analysis further reveals systematic failure propagation patterns and significant inefficiencies in resource utilization, with computational costs differing by up to three orders of magnitude (2525x) among comparably performing models. These findings suggest \ramp advances agentic model evaluation toward continuous, runtime-observable, and production-grounded assessment.
\end{abstract}


\maketitle

\section{Introduction}

Large language models (LLMs) have rapidly evolved from conversational assistants into autonomous software engineering agents capable of planning, implementing, debugging, and deploying increasingly complex software stacks and infrastructure workflows~\cite{jimenez2024swebench,liu2024agentbench,frontiereng2026}.
Recent agentic models have demonstrated strong capabilities in repository-level code generation, multi-file reasoning, iterative tool invocation, and workflow orchestration.
However, despite these advances, the evaluation methodologies for such systems remain largely rooted in a \emph{{\bf benchmark-driven paradigm}} consisting of isolated, static, and short-horizon tasks, where each evaluation instance is executed independently and the agent state is reset after every task.

This evaluation paradigm exhibits a fundamental mismatch with real-world production engineering environments.
In practice, \emph{{\bf production software and infrastructure engineering}} involve long-horizon iterative workflows spanning multi-file and cross-module reasoning, dependency management, runtime debugging, toolchain coordination, deployment validation, continuous integration pipelines. Real production systems must further operate under imperfect runtime conditions including environment instability, resource constraints, asynchronous execution, and cascading failure propagation.
Existing static benchmarks fail to capture these dynamics---they primarily measure instantaneous coding task completion rather than sustained engineering capability under evolving runtime conditions. Moreover, conventional benchmarks provide limited visibility into operational efficiency, execution stability, recovery behavior, and cost-performance tradeoffs, all of which are critical factors in practical deployment environments.

As a result, strong benchmark performance does not necessarily imply \underline{practical engineering utility}.
Models that perform well on benchmark-style coding may still fail in realistic production environments where runtime correctness, execution robustness, and adaptive recovery become critical.
In many cases, failures originate not from insufficient code generation ability, but from unstable tool interaction, environment inconsistency, cascading execution errors, and inefficient runtime orchestration. Existing benchmark paradigms largely overlook these system-level behaviors because they evaluate isolated task completion rather than continuous execution under evolving runtime conditions.
We therefore argue that the field must move \emph{beyond static benchmarks} toward production-grounded \emph{assessment workloads} to gauge agentic models in realistic, multi-stage engineering environments with persistent runtime state, controlled intermediate state injection, multi-dimensional metrics, and systematic failure diagnosis. Such assessment should not only measure whether a task succeeds, but also characterize how the system behaves throughout the execution process.

To this end, we present \ramp{} ({\it Runtime Assessment of Models in Production}), a production-grounded assessment infrastructure built upon the \yatcc{} integrated serving platform.
\ramp{} provides a unified evaluation architecture supporting heterogeneous LLM providers and agent SDKs, 
through standardized API access and orchestration services.
Unlike conventional benchmark pipelines, \ramp{} assesses models using realistic long-horizon compiler construction workloads organized as serial dependency chains, reflecting practical software engineering and infrastructure development processes, which enables systematic analysis of scaffolding effects and execution complexity under production-style workflows.

A key design of RAMP is the \resurrection mechanism, which decomposes cascading workflow failures into recoverable diagnostic stages through controlled intermediate-state restoration. This mechanism substantially improves runtime observability and enables fine-grained analysis of downstream execution behavior that would otherwise be hidden by early-stage failures. In addition, RAMP introduces utility-oriented multi-dimensional metrics that jointly assess outcome quality, execution efficiency, runtime cost, workflow robustness, recovery capability, and deployment-oriented operational behavior.

In summary, the paper makes the following contributions:
\begin{enumerate}[nosep]
\item We propose \ramp{} as a production-based assessment \emph{infrastructure}, not merely another benchmark, that moves beyond synthetic isolated tasks toward persistent, runtime-observable evaluation of agentic models under realistic engineering environments.
\item We construct realistic compiler construction workflows with strict serial dependencies, verifiable intermediate states that systematically quantify the impact of scaffolding and execution complexity in agentic software engineering tasks. Further, we introduce utility-oriented metrics that characterize not only task outcomes, but also runtime efficiency, failure patterns, and deployment-relevant operational cost.
\item We provide empirical evidence across \cnt model tests, demonstrating that realistic assessment reveals failure patterns invisible to static benchmarks and that resurrection-based recovery substantially improves downstream diagnostic observability.
\end{enumerate}
\section{Background and Motivation}

In this section, we first introduce \yatcc, detailing \yatcc Tasks and \yatcc Platform capabilities, and discuss how they reflect a production-level serving environment. We then analyze the motivation and challenges of evaluating agentic systems within such workflows, formalizing the key requirements that a realistic assessment framework should satisfy.

\subsection{\yatcc Tasks and Platform}
Here, we briefly introduce the practical compiler-construction tasks and the \yatcc support platform that together form the foundation of realistic model assessment.

\begin{table}[!t]
\centering
\caption{Compilation tasks of \yatcc. Each task consumes the output artifact of its predecessor, which can form an serial dependency.}
\small
\begin{tabular}{clll}
\toprule
\textbf{Task} & \textbf{Name} & \textbf{Objective} & \textbf{Input Artifact} \\
\midrule
\myt{0} & Env. Setup & Build \& verify toolchain & --- \\
\myt{1} & Lexer & Tokenize C source code & Preprocessed source \\
\myt{2} & Parser & Construct AST/ASG & Token stream (from \myt{1}) \\
\myt{3} & IR Gen. & Emit LLVM IR & AST/JSON (from \myt{2}) \\
\myt{4} & IR Opt. & Implement LLVM Pass & Raw IR (from \myt{3}) \\
\myt{5} & Asm. Gen. & Produce RV64 assembly & Optimized IR (from \myt{4}) \\
\bottomrule
\end{tabular}
\label{tab:tasks}
\end{table}

\subsubsection{Compiler-Construction Tasks}

The compiler-construction tasks built on \yatcc~\cite{yatcc2026} is based on LLVM~\cite{LLVMCGO04}, which has been extensively used as real-world compiler coursework and engineering exercises. The workload consists of six sequentially dependent compilation stages or tasks, where each task produces an intermediate artifact that becomes the input dependency for the next stage. Specifically, \myT{0} verifies agent capability and the build environment, \myT{1} produces a token stream, \myT{2} generates an AST/ASG, \myT{3} generates LLVM IR, \myT{4} performs IR optimization, and \myT{5} generates RV64 assembly. Each task is graded using dedicated test suites with comprehensive correctness checks, with scores ranging from 0 to 100.

\begin{figure}[t]
\centering
    \includegraphics[width=0.75\linewidth]{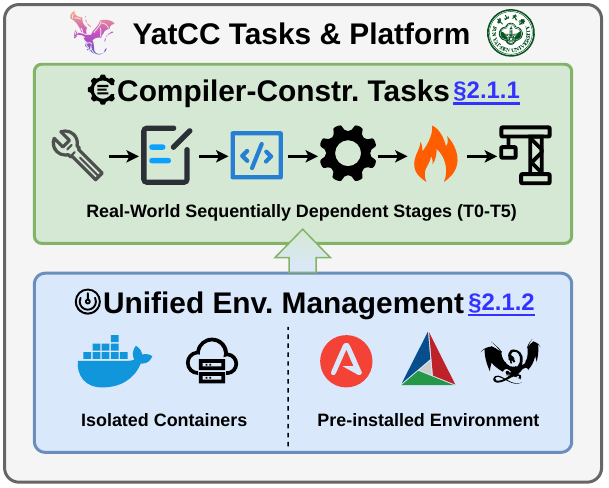}
\caption{\yatcc, with Compiler-Construction Tasks and Unified Environment Management.}
\label{fig:YatCC}
\end{figure}

Compiler development naturally embodies realistic systems-development jobs, involving large codebases, multi-stage execution pipelines, strict correctness constraints, and tightly coupled dependencies across lexical analysis, parsing, optimization, code generation, linking, and runtime validation. Each stage produces intermediate artifacts that must remain semantically and functionally consistent throughout the pipeline. Completing these tasks requires sustained interaction with external compilers, build systems, testing frameworks, and debugging tools, while continuously maintaining evolving project states and resolving execution failures. As a result, these tasks capture core characteristics of real-world software engineering, including long-horizon execution, iterative refinement, dependency management, and operational coordination in complex runtime environments.

\subsubsection{Unified Environment Management}

To reliably support these long-horizon compilation tasks and ensure strict dependency isolation, \yatcc Platform provides a robust execution infrastructure with unified environment management capabilities. With isolated containers and pre-installed environment, \yatcc Platform is a one-stop execution infrastructure that supports application development, deployment, execution, and management across heterogeneous execution environments through unified runtime abstractions. It integrates workflow runtime, containerized execution, external toolchains, and resource-managed infrastructure into a unified operational environment.

Applications running on \yatcc execute within complete software stacks involving dependency installation, compilation, testing, runtime monitoring, and iterative debugging, reflecting practical deployment and engineering workflows. Such environments naturally require continuous coordination across execution pipelines, system services, and runtime states over extended execution horizons. By leveraging unified containerized sandboxes and pre-installed toolchains, \yatcc Platform ensures that the full software stack and runtime configurations remain deterministic and isolated across repeated execution cycles, preventing environmental cross-contamination while tracking persistent state evolution.

\subsection{Motivation and Challenges}

The production-oriented characteristics of \yatcc and \yatcc Tasks expose key limitations in existing agent evaluation methodologies.

Overall, current evaluation approaches suffer from three major limitations. {\bf First}, they lack support for serial dependency and persistent execution state, preventing assessment of cross-task interactions, cumulative error propagation, and long-horizon workflow completion. {\bf Second}, they are largely outcome-centric, reporting only final scores without visibility into process efficiency, execution dynamics, failure behavior, or recovery capability. {\bf Third}, evaluation infrastructures are often tightly coupled to specific benchmarks or frameworks, making systematic cross-model and cross-agent comparison difficult.

\yatcc Tasks address these limitations by providing a unified, production-oriented evaluation setting. The workloads preserve execution continuity across sequential compilation stages, maintain evolving intermediate artifacts throughout the pipeline, and require sustained interaction with realistic build, testing, and debugging environments. These characteristics enable assessment not only of task completion, but also of execution robustness, dependency management, runtime coordination, and long-horizon operational behavior.

\begin{table*}[!t]
\centering
\caption{Benchmark comparison. \textcolor{deepgreen}{$\checkmark$} = supported, \textcolor{deepred}{$\times$} = not supported, \textcolor{deepblue}{partial} = partially supported. \emph{Real-world} = tasks derived natively from practical production systems; \emph{Exec.} = execution-grounded scoring; \emph{Chain Dep.} = tasks or actions form a sequential dependency chain; \emph{Strict Score} = zero-tolerance evaluation (e.g., non-perfect output yields zero); \emph{Artifact} = explicitly evaluates and preserves intermediate generated artifacts (e.g., AST, IR) across workflow stages; \emph{Process} = outputs detailed process metrics (e.g., cost, time) as first-class criteria; \emph{Eval Points} = the total number of distinct tasks or test instances evaluated.}
\small
\begin{tabular}{lccccccc}
\toprule
\textbf{Benchmark} & \textbf{Real-world} & \textbf{Exec.} & \textbf{Chain Dep.} & \textbf{Strict Score} & \textbf{Artifact} & \textbf{Process} & \textbf{Eval Points} \\
\midrule
SWE-bench~{\cite{jimenez2024swebench}} & \textcolor{deepgreen}{$\checkmark$} & \textcolor{deepgreen}{$\checkmark$} & \textcolor{deepred}{$\times$} & \textcolor{deepgreen}{$\checkmark$} & \textcolor{deepred}{$\times$} & \textcolor{deepred}{$\times$} & 2.3K \\
AgentBench~{\cite{liu2024agentbench}} & \textcolor{deepblue}{partial} & \textcolor{deepgreen}{$\checkmark$} & \textcolor{deepgreen}{$\checkmark$} & \textcolor{deepblue}{partial} & \textcolor{deepred}{$\times$} & \textcolor{deepred}{$\times$} & — \\
GAIA~{\cite{mialon2024gaia}} & \textcolor{deepblue}{partial} & \textcolor{deepred}{$\times$} & \textcolor{deepgreen}{$\checkmark$} & \textcolor{deepgreen}{$\checkmark$} & \textcolor{deepred}{$\times$} & \textcolor{deepred}{$\times$} & 466 \\
OSWorld~{\cite{xie2024osworld}} & \textcolor{deepgreen}{$\checkmark$} & \textcolor{deepgreen}{$\checkmark$} & \textcolor{deepgreen}{$\checkmark$} & \textcolor{deepblue}{partial} & \textcolor{deepred}{$\times$} & \textcolor{deepblue}{partial} & 369 \\
WebArena~{\cite{zhou2024webarena}} & \textcolor{deepgreen}{$\checkmark$} & \textcolor{deepgreen}{$\checkmark$} & \textcolor{deepgreen}{$\checkmark$} & \textcolor{deepgreen}{$\checkmark$} & \textcolor{deepred}{$\times$} & \textcolor{deepblue}{partial} & 812 \\
Terminal-Bench~{\cite{merrill2026terminalbench}} & \textcolor{deepgreen}{$\checkmark$} & \textcolor{deepgreen}{$\checkmark$} & \textcolor{deepgreen}{$\checkmark$} & \textcolor{deepgreen}{$\checkmark$} & \textcolor{deepred}{$\times$} & \textcolor{deepred}{$\times$} & 89 \\
SkillsBench~{\cite{li2026skillsbench}} & \textcolor{deepgreen}{$\checkmark$} & \textcolor{deepgreen}{$\checkmark$} & \textcolor{deepred}{$\times$} & \textcolor{deepgreen}{$\checkmark$} & \textcolor{deepred}{$\times$} & \textcolor{deepred}{$\times$} & 84 \\
RE-Bench~{\cite{rebench2024}} & \textcolor{deepgreen}{$\checkmark$} & \textcolor{deepgreen}{$\checkmark$} & \textcolor{deepgreen}{$\checkmark$} & \textcolor{deepred}{$\times$} & \textcolor{deepblue}{partial} & \textcolor{deepgreen}{$\checkmark$} & 7 \\
ProgramBench~{\cite{programbench2026}} & \textcolor{deepgreen}{$\checkmark$} & \textcolor{deepgreen}{$\checkmark$} & \textcolor{deepred}{$\times$} & \textcolor{deepgreen}{$\checkmark$} & \textcolor{deepred}{$\times$} & \textcolor{deepred}{$\times$} & 200 \\
\midrule
Desired & \textcolor{deepgreen}{$\checkmark$} & \textcolor{deepgreen}{$\checkmark$} & \textcolor{deepgreen}{$\checkmark$} & \textcolor{deepgreen}{$\checkmark$} & \textcolor{deepgreen}{$\checkmark$} & \textcolor{deepgreen}{$\checkmark$} & $\ge 100$ \\
\bottomrule
\end{tabular}
\label{tab:benchmark_comparison}
\end{table*}

Table~\ref{tab:benchmark_comparison} positions a desired assessment infrastructure against existing benchmarks across six key dimensions. Contemporary benchmarks demonstrate robust capabilities across various axes: {\tt SWE-bench}, {\tt ProgramBench}, and {\tt SkillsBench} excel in real-world execution-grounded tasks with strict zero-tolerance scoring, while frameworks like {\tt WebArena}, {\tt OSWorld}, and {\tt AgentBench} naturally embed chain dependencies through complex multi-step interactions (though some, like {\tt OSWorld}, permit partial rewards rather than binary evaluation). Furthermore, {\tt RE-Bench} pioneers the integration of deep process metrics (such as time and monetary cost) into the evaluation of long-horizon research tasks.

However, a critical gap remains in evaluating production-level engineering workflows. Existing benchmarks primarily evaluate end-to-end task success without explicit verification of intermediate states. In compiler-construction and similar industrial pipelines, the \emph{Artifact}---the intermediate product such as a parsed AST or compiled IR passed between dependent stages---is fundamentally crucial. The desired framework must simultaneously integrate real-world workflows, execution-grounded testing, strict serial dependencies, zero-tolerance scoring, continuous intermediate artifact verification, and detailed process-level metrics. Together, these properties enable a more granular evaluation of practical agent capabilities that are difficult to capture through conventional isolated-task or purely outcome-centric benchmarks. To effectively assess agentic models in production-oriented environments, several key requirements that must be met.

{\bf A unified runtime infrastructure for heterogeneous model assessment}.
Production environments typically involve diverse models, APIs, execution frameworks, workflow engines, and deployment backends. A practical assessment platform should therefore provide a unified infrastructure that enables reproducible and scalable evaluation across heterogeneous models and runtime systems. Such an infrastructure should support standardized service interfaces, workflow orchestration, environment provisioning, dependency isolation, and runtime management, allowing different models and execution frameworks to be evaluated under consistent operational conditions.
The infrastructure should also enable closed-loop assessment pipelines for automated testing, execution tracing, error diagnosis, result aggregation, and replayable experimentation, which helps better capture the evolving behavior of agentic systems under complex and changing production workloads.

{\bf Realistic workloads and diverse assessment perspectives}.
Conventional benchmarks often focus on isolated tasks with short execution horizons, which fail to capture the characteristics of real production systems. Production-grade assessment instead requires realistic workloads involving long-running workflows, external tool invocation, iterative execution feedback, environment interaction, failure recovery, and multi-stage task coordination. Moreover, assessment should support multiple evaluation perspectives, including functional correctness, execution efficiency, robustness under runtime perturbations, adaptability to dynamic environments, and operational scalability under concurrent workloads.

{\bf End-to-end observability and measurable runtime metrics}.
Assessing agentic systems in production settings requires comprehensive runtime visibility beyond task success rates alone. The assessment infrastructure should provide observable metrics spanning workflow execution, tool interaction, runtime behavior, resource consumption, latency characteristics, failure propagation, and execution stability. Such observability is essential for understanding not only whether a task succeeds, but also how the system behaves under realistic operational conditions, enabling deeper analysis of reliability, orchestration capability, and production readiness.

Together, these requirements highlight that \textbf{assessing agentic models in production systems fundamentally a systems-level problem rather than merely a benchmark-level evaluation task}. Effective assessment therefore necessitates integrated runtime infrastructure, realistic workload modeling, comprehensive observability, and scalable orchestration mechanisms.

\section{Design and Implementation}

To assess agentic models in production-level environments, we propose \ramp{}, which is a unified assessment infrastructure that cleanly decouples workload specification from execution, enabling scalable tests across heterogeneous model providers and agent frameworks through a standardized API layer. Unlike prior benchmark approaches, it treats both task outcomes and process dynamics as first-class analytical objects, allowing detailed, and fine-grained analysis of agentic model behavior. The system is structured around three core components—{\bf framework}, {\bf workloads}, and {\bf metrics}—providing a flexible yet systematic foundation for end-to-end assessment.

\subsection{Assessment Framework}

\begin{figure}[h]
\centering
    \includegraphics[width=\linewidth]{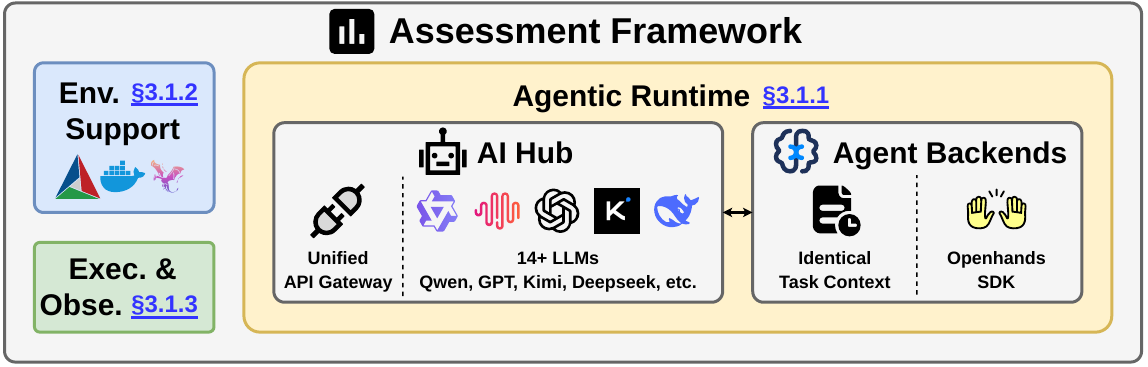}
\caption{The framework of \ramp{} that enables unified and extensible integration of heterogeneous agentic models and backends.}
\label{fig:design}
\end{figure}

The assessment framework is organized into three layers of runtime, environment and execution, which together provide a unified, reproducible, and extensible foundation for assessing LLM agents in production-oriented software engineering workflows.

\subsubsection{Agentic Runtime Layer.}
\label{sec:Agentic Runtime Layer}
This layer provides a unified abstraction over heterogeneous agent frameworks and model providers. It supports multiple agent backends, each representing a distinct interaction paradigm. 
All backends consume a shared task context format and generate standardized execution outputs, thus enabling direct cross-framework comparisons.

To further normalize model access, \ramp{} integrates a unified API gateway, i.e., AI Hub~\footnote{AI Hub: \url{https://aihub.arcsysu.cn}} that exposes OpenAI-compatible interfaces \cite{openai2026api} for both proprietary and open-weight models. This design eliminates backend-specific integration overhead and ensures that observed assessment differences primarily reflect model and scaffold capabilities rather than infrastructure inconsistencies.

\subsubsection{Environment Support Layer.}
This layer ensures execution isolation and reproducibility. Each assessment run is executed within an independent workspace containing a complete copy of the target repository with all required dependencies pre-installed, including {\it ANTLR}, {\it LLVM 14}, and {\it pybind11}. In addition, every run launches a fresh containerized environment that is destroyed after execution, preventing cross-run contamination.

The build and execution environment is configured using a fixed CMake-based toolchain, and all scoring scripts are deterministic. This setup guarantees that assessment results remain stable and comparable across repeated runs and heterogeneous agent backends.

\begin{figure*}[!t]
    \centering
    \includegraphics[width=0.8\textwidth]{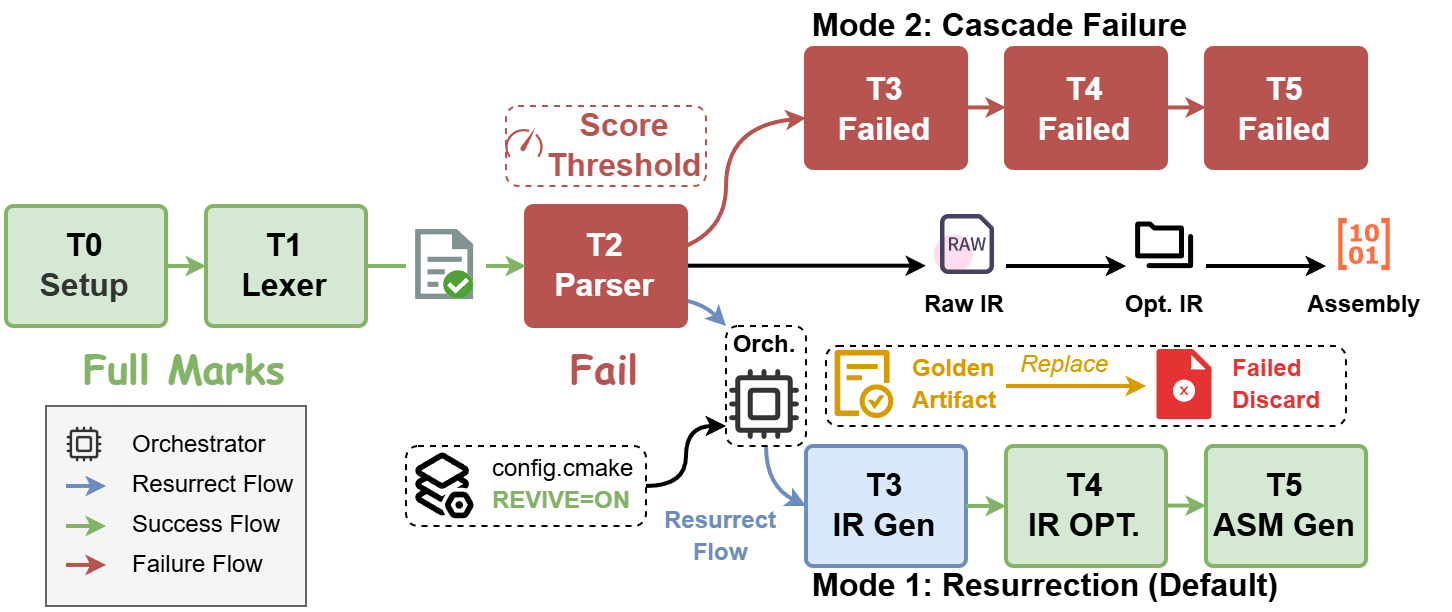}
    \caption{Long-horizon assessment workloads in the integrated pipeline of \ramp{} (\S\ref{long-horizon-workloads}). The workflow demonstrates Serial Evolution where tasks pass cumulative context. When Task 2 fails (red), Mode 2 (wavy orange line) propagates the error, leading to cascading downstream failures. Mode 1 (solid blue line) transparently injects a golden artifact via the orchestrator, isolating the failure and allowing evaluation to continue.}
    \label{fig:integrated_pipeline}
\end{figure*}

\subsubsection{Execution and Observation Layer.}
The execution layer manages workload orchestration, execution monitoring, and result analysis. It is responsible for test scheduling, dependency resolution, resurrection triggering, and assessment pipeline coordination. The core component of this layer, referred to as the {\bf orchestrator}, maintains the state of serial task chains, tracks per-task pass/fail outcomes, and automatically triggers resurrection when task completion falls below the predefined threshold (e.g., 60\%). 

Beyond execution management, this layer also collects and aggregates assessment artifacts, including scores, logs, trajectories, and process metrics. It computes utility-oriented metrics, maintains dynamic leaderboards, and preserves all raw assessment data for reproducibility and fine-grained behavioral analysis.

\subsection{Long-horizon Workloads}
\label{long-horizon-workloads}

\ramp{} assesses LLM agents using long-horizon, production-oriented software engineering workloads grounded in compiler construction, as illustrated in Figure~\ref{fig:integrated_pipeline}. Unlike isolated coding benchmarks, \ramp{} organizes tasks into a serial dependency chain that preserves execution context, intermediate artifacts, and runtime state across the entire workflow. To further improve diagnostic resolution in long execution chains, \ramp{} introduces a resurrection protocol that enables continued assessment after intermediate failures.

\subsubsection{Serial Evolution}
\ramp{} currently provides a representative compiler-construction workload derived from the individual tasks~\cite{yatcc2024}. The workload consists of six sequential tasks spanning the compiler development pipeline. Each task provides partially implemented scaffold code, requiring agents to complete missing functionality within an existing codebase rather than generating standalone solutions from scratch.

A key characteristic of the workload is persistent execution continuity. The full repository state, execution context, and runtime environment are preserved across all tasks, simulating realistic agentic software development workflows in which agents incrementally evolve a single system over time. This design avoids artificial context resets between tasks and enables evaluation of continuous development capabilities, long-horizon dependency handling, and cumulative error propagation.

\subsubsection{Resurrection Protocol}
Serial workloads introduce a fundamental assessment challenge: failure at Task~$k$ may invalidate all downstream tasks, making it impossible to determine whether subsequent failures are caused by limited downstream capability or simply inherited broken states. To address this issue, \ramp{} introduces the resurrection protocol.

When a task score falls below the passing threshold (e.g., 60\%), resurrection is automatically triggered by the orchestrator. The failed intermediate artifact is replaced with the corresponding golden artifact generated by the reference implementation, and the build environment is reconfigured through \texttt{config.cmake} \\(\texttt{TASK\_k\_REVIVE=ON}). The agent then continues execution from Task~$k+1$ using the corrected system state.

Importantly, \resurrection is designed as an assessment protocol rather than a scoring adjustment mechanism. Its purpose is to separate two otherwise entangled failure modes: (1) the inability to correctly complete the current task, and (2) the capability to solve downstream tasks given valid prerequisites. This decomposition enables node-level capability analysis even when the overall pipeline is broken, substantially improving diagnostic granularity for long-horizon assessment.

To avoid behavioral bias, agents are not informed when resurrection occurs. All interventions are handled externally by the task orchestrator.

\subsubsection{Execution Modes}\label{section:Execution Modes}
\ramp{} supports two execution modes for long-horizon workloads.

{\bf Mode 1: Serial Pipeline with Resurrection (Default)}.
Agents execute Tasks~0–5 sequentially. Whenever a task fails, resurrection is triggered and execution continues using the corrected intermediate artifact. This setting maximizes diagnostic coverage and produces the primary leaderboard containing both outcome and process metrics.

{\bf Mode 2: Serial Pipeline without Resurrection (Cascade failure)}.
Agents also execute Tasks~0–5 sequentially, but no resurrection is applied. Failed intermediate artifacts are directly propagated to downstream tasks, often causing cascading failures. This setting measures \emph{zero-shot pipeline depth}, i.e., how far an agent can progress in a fully autonomous end-to-end workflow without external correction.

Comparing these two settings allows \ramp{} to quantify the extent to which resurrection recovers diagnostic signals that would otherwise be obscured by cascading failures in long-horizon software engineering workflows.

\subsection{Multi-dimensional Metrics}
\label{sec:multi-dimension}
\ramp{} adopts a multi-dimensional measuring suite that moves beyond single accuracy-based scores toward utility-oriented assessment. In long-horizon software engineering workflows, successful assessment requires measuring not only whether an agent completes tasks, but also how efficiently it operates, how resources are consumed, and how failures emerge during execution. To this end, \ramp{} combines outcome metrics, process metrics, failure analysis, and composite utility measurements into the unified assessment framework.

\subsubsection{Outcome and Process Metrics}
\textbf{Outcome metrics.}
\ramp{} defines \emph{Mean Reward (MR)} as the primary outcome metric. MR computes a weighted average of task scores while incorporating resurrection-aware bonuses:

$$\text{MR} = \sum s_i w_i b_i / \sum w_i b_i$$,

where $s_i$ denotes the task score, $w_i = [0.05, 0.20, 0.20, 0.15, 0.30, 0.10]$ represents task importance weights, and $b_i$ is a resurrection factor. Tasks completed without resurrection receive a bonus ($b_i=1.2$), while resurrected tasks use the default weight ($b_i=1.0$). This formulation rewards both correctness and autonomous pipeline continuity.

\textbf{Process metrics.}
Beyond final task outcomes, \ramp{} records detailed execution dynamics throughout the assessment process. Collected metrics include token consumption, interaction turns, command executions, retry counts, and elapsed wall-clock time at both per-task and end-to-end levels. These measurements enable fine-grained analysis of agent efficiency, interaction behavior, and resource utilization.

\subsubsection{Failure Taxonomy}
Outcome scores alone provide limited insight into why agents fail in long-horizon workflows. To improve diagnostic interpretability, \ramp{} introduces an observable failure taxonomy grounded in and extended from the four empirically validated architectural fault dimensions proposed by Shah et al.\cite{shah2026characterizingfaultsagenticai}. We add a dedicated planning dimension to comprehensively cover all execution failure modes observed in long-horizon tasks. All categories are defined based on trace-observable behaviors, avoiding ambiguous cognitive labels:
\begin{itemize}
    \item \textbf{Reasoning  Failure:} Failures manifesting as repetitive debugging loops, infinite search cycles, or stalled iterations with no meaningful progress, arising from flawed control orchestration and execution loop logic.
    \item \textbf{Planning Failure:} Failures where the agent explicitly decides to bypass one or more required task steps, resulting from incorrect task decomposition or flawed strategic decision-making.
    \item \textbf{Context Failure:} Failures occurring when accumulated prompts, execution history, or intermediate state exceed the model's context window capacity, preventing further reasoning or action.
    \item \textbf{Tooling \& Integration Failure:} Failures caused by invalid tool invocations, external service access errors, resource interaction problems, or network connectivity issues.
    \item \textbf{Infrastructure Failure:} Failures stemming from bugs or design flaws in the agent framework itself, including dependency conflicts, environment misconfigurations, and broken exception handling.

\end{itemize}

\begin{table*}[!t]
\centering
\caption{Categorization and Abbreviations of the Evaluated Models. Model types: FF (Frontier Flagship), AM (Advanced Mainstream), SL (Standard \& Lightweight).}
\label{tab:model_groups}
\small
\begin{tabular}{lll|lll|lll}
\toprule
\textbf{Model} & \textbf{Type} & \textbf{Abbr.} & \textbf{Model} & \textbf{Type} & \textbf{Abbr.} & \textbf{Model} & \textbf{Type} & \textbf{Abbr.} \\
\midrule
claude-opus-4-7     & FF & Opus-4.7     & kimi-k2.6         & FF & Kimi-2.6     & kimi-k2.5         & AM & Kimi-2.5    \\
gpt-5.5             & FF & GPT-5.5      & deepseek-v4-flash & AM & DS-v4-Flash  & glm-4.6           & SL & GLM-4.6     \\
deepseek-v4-pro     & FF & DS-v4-Pro    & deepseek-reasoner & AM & DS-Reasoner  & deepseek-chat     & SL & DS-Chat     \\
qwen3.6-max-preview & FF & Qwen-3.6-Max & qwen3.5-plus      & AM & Qwen-3.5-Plus& qwen3-coder-flash & SL & Qwen3-Coder \\
glm-5.1             & FF & GLM-5.1      & glm-4.7           & AM & GLM-4.7      & minimax-m2.5      & SL & MiniMax-2.5 \\
\bottomrule
\end{tabular}
\end{table*}

\subsubsection{Overall Utility}
\label{sec:aei}
The preceding metrics independently characterize task completion, execution efficiency, and failure behavior. However, practical deployment scenarios often require assessing overall engineering productivity under constrained resources. To capture this tradeoff, \ramp{} introduces the \emph{Agent Efficiency Index (AEI)}, a composite utility metric that jointly measures task effectiveness and resource efficiency:

\[
\mathrm{AEI}
=
\frac{1}{|\mathcal{D}|}
\sum_{d \in \mathcal{D}} s_d,
\qquad
\mathcal{D}
=
\{\mathrm{stage}, \mathrm{reward}, \mathrm{time}, \mathrm{cost}, \mathrm{tokens}\}.
\]
where $s_d \in [0,100]$ is the normalized score for dimension $d \in \mathcal{D}$.
Each dimension is mapped using the maximum value observed across all evaluated models.
Let $S$ denote the furthest pipeline stage reached, $R$ the mean reward, and $T$, $C$, and $K$ the end-to-end wall-clock time, LLM API cost (USD), and total token usage, respectively.
Let $S_{\max}$, $R_{\max}$, $T_{\max}$, $C_{\max}$, and $K_{\max}$ denote these observed maxima.
We define

\begin{align*}
& s_{\mathrm{stage}}  &&= \frac{S}{S_{\max}}\times 100,
& \quad & s_{\mathrm{reward}}  &&= \frac{R}{R_{\max}}\times 100, \nonumber\\
& s_{\mathrm{time}}   &&= \frac{T_{\max} - T}{T_{\max}}\times 100,
& \quad & s_{\mathrm{cost}}    &&= \frac{C_{\max} - C}{C_{\max}}\times 100, \nonumber\\
& s_{\mathrm{tokens}} &&= \frac{K_{\max} - K}{K_{\max}}\times 100.
\end{align*}

Because each raw metric is non-negative and at most its observed maximum, every $s_d$ lies in $[0,100]$ by construction. And,
\emph{stage} captures pipeline depth (the furthest stage completed along the serial workflow),
\emph{reward} captures \emph{Mean Reward (MR)},
and \emph{time}, \emph{cost}, and \emph{tokens} capture elapsed time, monetary API cost, and token consumption, respectively.

Higher \aei values indicate more balanced utility across outcome quality and resource utilization.
Because each dimension is equally weighted, \aei cannot be maximized through strong task scores alone while ignoring time, cost, or context pressure.

\section{Results and Analysis}
In this section, we first describe the experimental methodology, including the tested models, execution settings, and assessment metrics. We then present and analyze the experimental results across multiple dimensions.

\subsection{Experimental Methodology}
\subsubsection{Models and Agent Backends}
We evaluate RAMP using \cnt representative models spanning both proprietary and open-weight ones, as listed in Table~\ref{tab:model_groups}. The assessed models cover multiple capability tiers, including frontier flagship models, advanced mainstream models, and lightweight efficient models, enabling comprehensive analysis with varying performance–cost tradeoffs.

All assessments are conducted using a unified agent backend~\footnote{Results of extra backends are to be added in analysis.} built on top of the OpenHands SDK~\cite{wang2026openhandssoftwareagentsdk}, enabling agents to iteratively invoke tools, inspect runtime outputs, modify source codes, and coordinate execution workflows within a persistent runtime environment.
To ensure fair comparison across models, all LLM services are accessed through the AIHub of \ramp{} (\S\ref{sec:Agentic Runtime Layer}), which exposes standardized OpenAI-compatible interfaces for heterogeneous model providers.
For each model test, agents execute the complete serial workload under identical runtime conditions, using the same execution environment, toolchain configuration, prompts, and orchestration policies.

\subsubsection{Configurations and Metrics}
All experiments are performed under a standardized assessment pipeline. For each run, the agent is initialized with a fresh workspace containing the complete repository, pre-installed dependencies, build toolchains, and grading scripts. Agents interact with the environment through iterative command execution, file modification, compilation, testing, and debugging operations until task completion or termination.

During execution, \ramp{} continuously records both outcome and process traces, including task scores, interaction histories, command invocations, execution logs, token consumption, runtime latency, and resource usage statistics. All raw artifacts are preserved to support reproducibility and detailed post-hoc behavioral analysis.

All experiments are executed on a standardized hardware and software platform to ensure reproducibility and fair comparison across evaluated agents. The evaluation server is configured with with following settings: \textcircled{1} CPU: AMD EPYC 7742 64-Core Processor; \textcircled{2} Memory: 256 GB RAM; \textcircled{3} Operating System: Debian GNU/Linux 11 (bullseye); \textcircled{4} Libraries and Toolchains: LLVM 14, ANTLR, pybind11, CMake, Docker, and related runtime dependencies.
And, to bound evaluation cost while preserving sufficient exploration capability, we enforce a strict maximum of 500 dialogue turns per agent across the full workload pipeline.

We further adopt a multi-dimensional metric suite that measures not only task correctness, but also execution efficiency, failure behavior, and overall engineering utility. As defined in \S\ref{sec:multi-dimension}, the evaluation consists of four complementary dimensions:
\begin{itemize}
\item {\bf Performance}: measured using capability depth, pipeline completion, and weighted leaderboard scores;
\item {\bf Cost}: including wall-clock execution time, token consumption, command counts, and estimated monetary cost;
\item {\bf Failure}: characterized through observable failure categories such as execution failure, process collapse, cost overrun, task skipping, and context exhaustion;
\item {\bf Overall Utility}: measured using the Agent Efficiency Index (AEI), a composite metric that jointly captures task utility and resource efficiency(\S~\ref{sec:aei}).
\end{itemize}

Together, these metrics provide a holistic characterization of agent behavior under realistic long-horizon software engineering workloads.

\subsection{Capability Boundaries and Leaderboard Scoring}

\begin{table*}[!t]
\centering
\caption{\ramp{} leaderboard. MR = Mean Reward. \textbf{Bold} values indicate task scores $\geq 60$. The \textsc{Baseline} row corresponds to the unmodified code framework without agent intervention.}
\small
\begin{tabular}{clrrrrrrrrr}
\toprule
\textbf{\#} & \textbf{Model} & \textbf{T0} & \textbf{T1} & \textbf{T2} & \textbf{T3} & \textbf{T4} & \textbf{T5}  & \textbf{MR} & \textbf{Cost\$} \\
\midrule
\textbf{1} & Opus-4.7  & \textbf{100} & \textbf{100} & \textbf{100} & \textbf{100.0} & \textbf{68.4} & \textbf{100}  & 93.39 & 126.24 \\
\textbf{2} & DS-v4-Pro & \textbf{100} & \textbf{100} & \textbf{100} & \textbf{83.6} & 38.6 & \textbf{100}  & 85.34 & 8.68 \\
\textbf{3} & DS-v4-Flash& \textbf{100} & \textbf{100} & 31.6 & 37.0 & 39.0 & \textbf{100}  & 66.48 & 3.43 \\
\textbf{4} & GPT-5.5 & \textbf{100} & \textbf{100} & \textbf{100} & \textbf{100} & 48.2 & 0.0  & 65.91 & 8.77 \\
\textbf{5} & Qwen-3.6-Max & \textbf{100} & \textbf{100} & \textbf{100} & \textbf{65.8} & 35.0 & 0.0 & 56.72 & 67.28 \\
\textbf{6} & Kimi-2.5 & \textbf{100} & \textbf{100} & 14.2 & 24.7 & 39.7 & 9.4  & 40.24 & 1.75 \\

\textbf{7} & Kimi-2.6 & \textbf{100} & \textbf{95.9} & 31.4 & 1.4 & 37.5 & 0.0  & 33.94 & 3.99 \\

\textbf{8} & GLM-4.6 & \textbf{100} & \textbf{67.4} & 38.5 & 1.4 & 39.6 & 0.0  & 30.88 & --- \\
\textbf{9} & DS-Chat & \textbf{100} & \textbf{72.9} & 30.9 & 0.0 & 35.7 & 0.0  & 29.74 & 0.07 \\
\textbf{10} & Qwen-3.5-Plus & \textbf{100} & \textbf{100} & 0.0 & 0.0 & 38.0 & 0.0  & 29.63 & 14.37 \\
\textbf{11} & GLM-4.7 & \textbf{100} & \textbf{94.8} & 0.0 & 1.4 & 36.2 & 0.0  & 29.50 & 0.80 \\
\textbf{12} & GLM-5.1 & \textbf{100} & 0.0 & \textbf{97.7} & 0.0 & 36.6 & 0.0  & 27.26 & 2.03 \\
\textbf{13} & Qwen3-Coder & \textbf{100} & 56.0 & 20.6 & 1.4 & 35.1 & 0.0  & 25.97 & 0.05 \\
\textbf{14} & MiniMax-2.5 & \textbf{100} & \textbf{64.2} & 0.0 & 2.7 & 36.4 & 0.0  & 25.09 & 0.17 \\

\textbf{15} & DS-Reasoner & \textbf{100} & 0.0 & 0.0 & 0.0 & 0.0 & 0.0  & 7.69 & 0.23 \\
\midrule
\textbf{-} & \textsc{Baseline} & 100 & 17.6 & 20.6 & 1.4 & 35.1 & 0.0  & 23.38 & --- \\
\end{tabular}
\label{tab:main}
\end{table*}

To establish a rigorous baseline to measure agent capabilities, we conduct all \cnt model test runs under the most demanding assessment conditions, i.e., Mode~2 (serial pipeline without resurrection, \S\ref{section:Execution Modes}). This long-horizon regime eliminates all external intervention and enforces production-grade functional correctness, strictly respecting the dependency structure of real-world development—agents cannot proceed to a task until all preceding stages are fully and correctly completed.

Results reported in Table~\ref{tab:main} reveal a stark limitation: even the state-of-the-art models, including {\tt Opus-4.7} and {\tt GPT-5.5}, fail to complete the entire six-task pipeline. The vast majority of models suffer catastrophic early-stage malfunctions, and even the top-performing model {\tt Opus-4.7}
stalls at the IR Generation stage (\myt{3}), unable to fully complete IR Optimization (\myt{4}). Only a small subset of proprietary models like {\tt DS-v4-Pro} and {\tt Qwen-3.6-Max} reach \myt{3} with none progressing further. This demonstrates a clear capability ceiling for current models on complex multi-step reasoning pipelines.

\begin{figure*}[t]
\centering
\includegraphics[width=\linewidth]{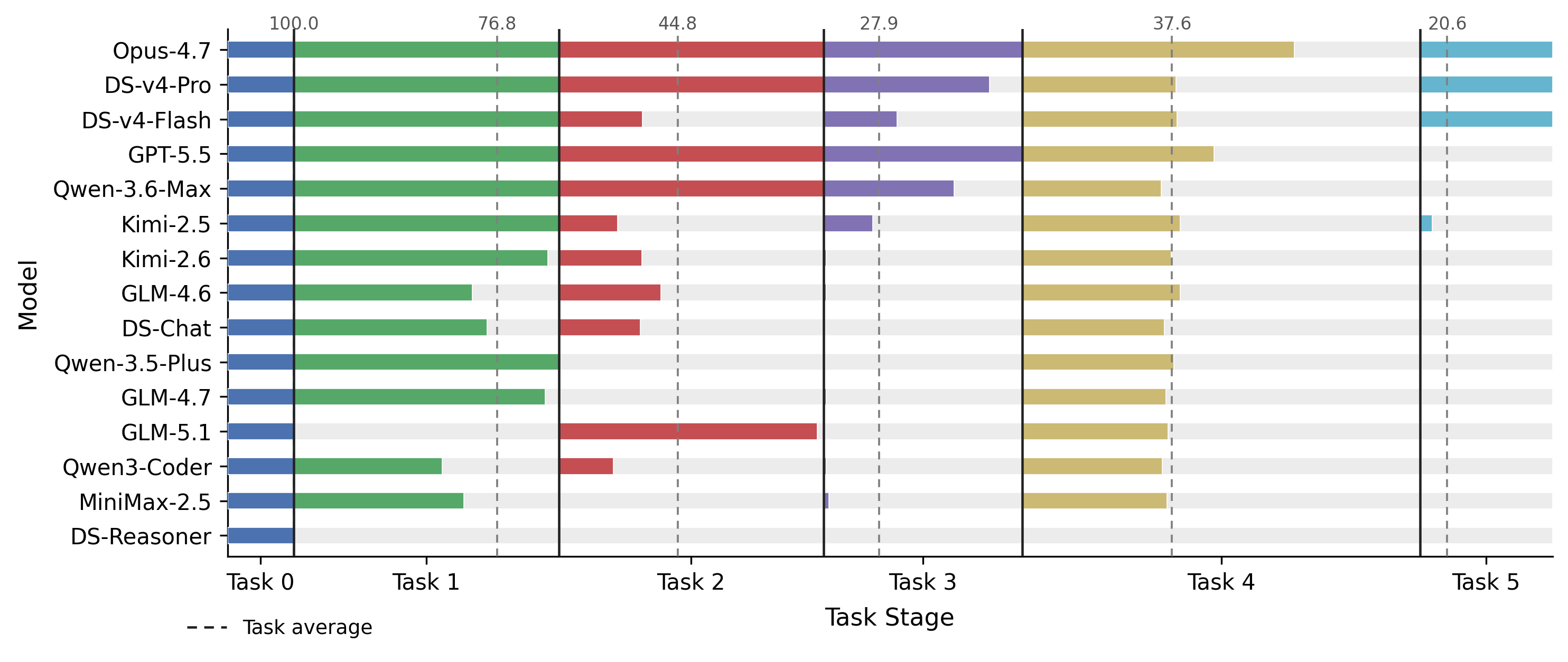}
\caption{Per-task score distribution of LLM agents. Models (y-axis) are ranked by overall mean reward. The x-axis is divided into six equal segments corresponding to \myT{0}–\myT{5}: environment setup, lexer, parser, IR generation, IR optimization, and backend code generation. Within each segment, the colored bar length indicates the normalized task score (out of 100).}
\label{fig:trajectory}
\end{figure*}

Table~\ref{tab:main} and Figure~\ref{fig:trajectory} together present the detailed \ramp{} leaderboard, moving beyond binary pass or fail boundaries to continuous per-task scoring. We summarize the key findings below.

\textbf{Finding 1: Current agentic models lack sustained multi-stage engineering capabilities for end-to-end autonomous development of complex engineering systems.} None of the \cnt assessed models achieve perfect scores across all six stages, confirming that flawless compiler chain implementation exceeds current zero-shot LLM agent capabilities even with structured scaffolding. {\tt Opus-4.7} leads with an MR of 93.39, achieving perfect completion on five out of six tasks but falling short at IR optimization (\myt{4}, 68.4). {\tt Deepseek-v4-pro} ranks second (MR 85.34) with four perfect tasks, while {\tt GPT-5.5} also achieves four perfect tasks but produces non-functional output at backend code generation (\myt{5}, 0.0).

Task completion rates across stages are 100\% (\myt{0}), 46.7\% (\myt{1}), 26.7\% (\myt{2}), 13.3\% (\myt{3}), 0\% (\myt{4}), and 20\% (\myt{5}). Notably, \myt{5} exhibits a higher task completion rate than \myt{4}. This means task difficulty itself does not follow a monotonic increasing pattern, so the observed performance decline cannot be simply attributed to harder tasks. Instead, performance declines reflect the cumulative impact of earlier task modifications to the execution environment and the propagation of subtle errors across stages. The overall average MR of 42.7 is only 17.6 points above the non-agent {\tt Baseline}, highlighting that current agents provide limited resilience to state accumulation and error propagation in multi-stage engineering workflows.

\textbf{Finding 2: Agentic models show systematic capability decline in long-horizon execution.}
All models achieve perfect scores on \myt{0} (environment setup), and 8 models achieve perfect scores on \myt{1} (lexer). However, performance declines sharply at \myt{2} (parser) and \myt{3} (IR generation) as the execution environment becomes increasingly modified and interdependent. Models with perfect \myt{1} scores experience an average 41.3-point score lowering by \myt{3}, with one model (Qwen-3.5-Plus) achieving scores of 0 points. Quantitatively, the average stage-wise score decreases sequentially from 100 (\myt{0}) to 76.75 (\myt{1}), 44.85 (\myt{2}), 27.95 (\myt{3}), 37.60 (\myt{4}), and 20.63 (\myt{5}), confirming a consistent downward trend despite non-monotonic fluctuations.

Three distinct performance patterns emerge across the pipeline. First, gradual degradation is observed in several models like Kimi-2.6 and DS-Chat, where scores decrease incrementally across consecutive stages. Second, error amplification occurs when minor imperfections in earlier stages lead to disproportionately lower scores in subsequent stages. This is observed in {\tt DS-v4-Pro}, which achieves 83.6 on \myt{3} but only 38.6 on \myt{4}. Third, discontinuous performance manifests as unexpected score variations in later stages that are uncorrelated with earlier performance. Most notably, {\tt GPT-5.5} achieves perfect scores on \myt{0}–\myt{3} but produces completely non-functional backend code, while {\tt DS-v4-Flash} exhibits limited performance at \myt{2} and \myt{3} but delivers perfect \myt{5} output. This pattern demonstrates that current LLMs exhibit inconsistent performance across stages that share dependencies, with distinct capability profiles: for example, {\tt Opus-4.7} excels in IR optimization while {\tt DS-v4-Pro} shows superior backend code generation ability.

\subsection{Process Efficiency}
We next move beyond performance analysis to examine the efficiency implications, including execution time, token consumption, and monetary cost.

The assessment reveals extreme variability in cost efficiency, with total inference costs ranging from \$0.05 ({\tt Qwen3-Coder}) to \$126.24 ({\tt Opus-4.7})—a 2,525x difference. API costs are calculated based on provider-specific pricing for input and output tokens separately. {\tt Opus-4.7} delivers the highest mean reward (MR=93.39) but at a prohibitive cost, representing a 14.5x cost premium over {\tt DS-v4-Pro} (MR=85.34, \$8.68) for only a 9.4\% improvement on overall performance. {\tt DS-v4-Pro} achieves the best balance of performance and cost, delivering 29.5\% higher MR than {\tt GPT-5.5} (MR=65.91, \$8.77) at nearly identical cost. Most strikingly, {\tt Qwen-3.6-Max} ranks fifth in performance (MR=56.72) but incurs 7.75x higher cost than {\tt DS-v4-Pro} and 19.6x higher cost than {\tt DS-v4-Flash} (MR=66.48, \$3.43). At the lower end, {\tt DS-Chat} (\$0.07) and {\tt Qwen3-Coder} (\$0.05) outperform the baseline by 27.2\% and 11.1\% respectively at negligible cost, making them attractive options for rapid prototyping.

\begin{figure*}[!t]
\centering
\includegraphics[width=0.95\textwidth]{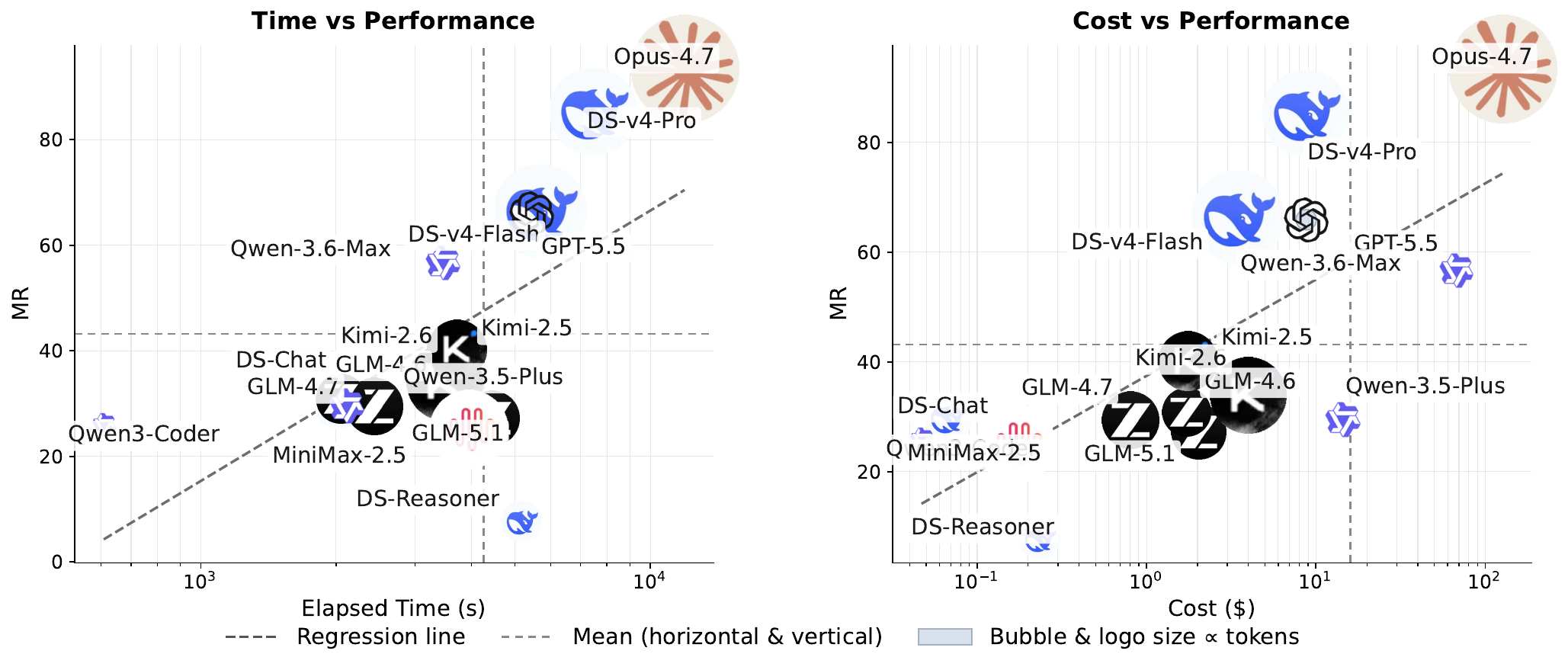}
\caption{Trade-off of cost and performance: elapsed time (left) and API cost (right) vs.\ mean reward across \cnt models ($n{=}15$). Dashed lines are OLS fits (linear fitted lines) of mean reward on $\log_{10}(x)$ (log-scaled axes). Moderate positive association: $R^2 \approx 0.40$ (time) and $R^2 \approx 0.52$ (cost). Bubble area is proportional to total LLM tokens.}
\label{fig:cost}
\end{figure*}

Figure~\ref{fig:cost} plots the relationship between computational investment (elapsed time and LLM API cost) against outcome quality.

\textbf{Finding 3: Agentic models exhibit substantial variability in performance relative to cost and execution time.}
On the log-scaled axes, ordinary least squares (OLS), a widely used linear regression technique,indicates a moderate positive association between computational investment and performance: API cost against mean reward yields $R^2 \approx 0.52$, while elapsed time shows a somewhat weaker but still visible trend with $R^2 \approx 0.40$. As the goodness-of-fit metric of OLS, $R^2$
 measures the proportion of performance variance explained by resource overhead, with values closer to 1 indicating a stronger linear correlation. These results suggest that greater resource consumption is \emph{associated} with higher reward on average, but the relationship is far from deterministic and leaves substantial variance unexplained. In particular, several high-cost or long-running systems still achieve only moderate performance, whereas some more economical configurations remain highly competitive. For example, {\tt Qwen-3.6-Max} is among the most expensive models yet attains only moderate reward, while {\tt DS-v4-Flash} reaches top-tier performance at substantially lower cost. Overall, computational budget and execution time provide useful but incomplete signals of agent quality: they correlate with performance yet cannot reliably predict it.

\subsection{Failure Analysis}

\begin{figure*}[t]
\centering
\includegraphics[width=0.9\linewidth]{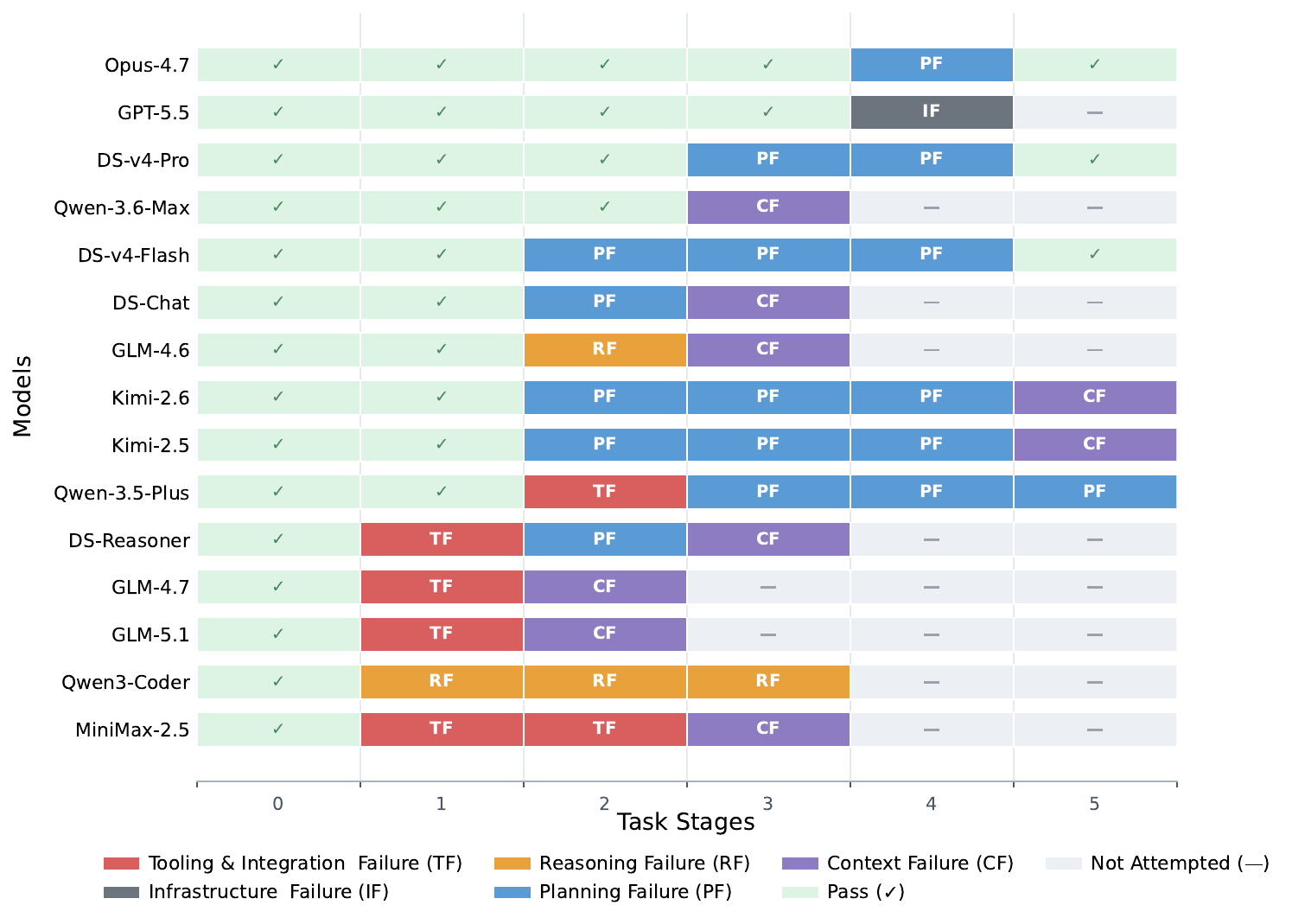}

\caption{Multi-dimensional efficiency profiles. The radar axes show pipeline stage, mean stage-wise reward, inverted wall-clock time, inverted LLM cost, and inverted token usage. Each axis is mapped to $[0,100]$ using the leaderboard-wide normalization in \S\ref{sec:aei}. Larger values indicate stronger progress, higher reward, or lower resource use relative to the costliest run among evaluated models.}

\label{fig:failure_taxonomy}
\end{figure*}

Existing benchmarks rarely systematically categorize runtime failure modes in long-horizon serial workflows. By contrast, RAMP classifies agent failures into five mutually exclusive trace-driven categories of {\it Reasoning Failure}, {\it Planning Failure}, {\it Context Failure}, {\it Tooling \& Integration Failure}, and {\it Infrastructure Failure}. Each category follows explicit observable criteria derived from execution logs and interaction traces to avoid subjective interpretation, and only the primary root cause is labeled when multiple failure symptoms co-occur.
Figure~\ref{fig:failure_taxonomy} summarizes the failure taxonomy across all evaluated models.

Statistically, {\it Context Failure} emerges as the most prevalent hard-stop failure across the \cnt evaluated models (9 out of \cnt, 60.0\%), followed by {\it  Planning Failure} (2 out of \cnt, 13.3\%) and {\it Reasoning Failure} (1 out of \cnt, 6.7\%). Further stage-distributed analysis reveals clear failure concentration patterns: {\it Context Failure} predominantly occurs at \myt{2}–\myt{3}, where accumulated code artifacts, dialogue history, and task instructions continuously consume context budget. {\it  Planning Failure} mainly appears at \myt{2}–\myt{4}, while {\it Tooling \& Integration Failure} is concentrated in \myt{2} (parser) and \myt{3} (IR generation) stages that require frequent tool invocation and environment interaction.

\textbf{Finding 4: Long-horizon execution of current agents is primarily bottlenecked by context exhaustion and premature strategic abandonment.}
Context Failure stems from limited context window capacity and ineffective long-range context compression, rather than fundamental reasoning deficiencies. It can be precisely identified via API logs and trajectory truncation signals, acting as a critical system-level constraint for sustained agent workflow execution.

{\it Planning Failure} can be observed in 8 of the \cnt evaluated models (53.3\%) and act as the primary termination cause in 2 models (13.3\%), making it one of the most common observed failure behaviors. Unlike other failures, {\it Planning Failure} originates from explicit agent strategic decisions, where the agent acknowledges unresolved task requirements but voluntarily skips difficult stages to conserve budget or reduce iteration pressure. This behavior represents premature task abandonment rather than adaptive scheduling, as the agent redefines task boundaries instead of completing the predefined serial workflow.

\begin{figure*}[t]
    \centering
    \includegraphics[width=\textwidth]{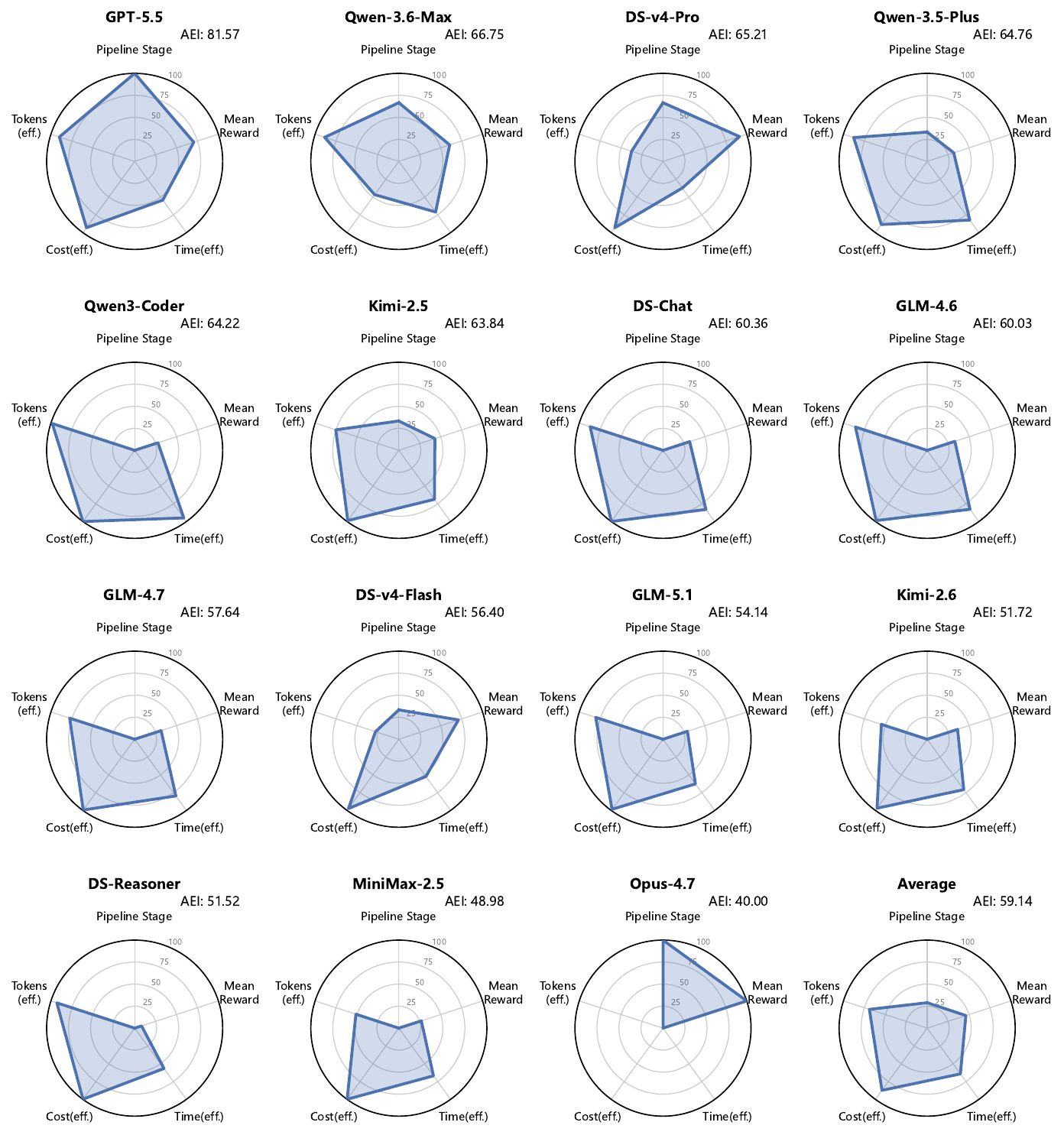}
    \caption{Multi-dimensional efficiency profiles. The radar axes show pipeline stage, mean stage-wise reward, inverted wall-clock time, inverted LLM cost, and inverted token usage, each mapped to $[0,100]$ using fixed reference scales. Larger values therefore indicate stronger progress, higher reward, shorter runtime, lower cost, or lower token consumption under the same external budget.}
    \label{fig:fig_radar}
\end{figure*}

Further behavioral abstraction derived from {\it Planning Failure} reveals two representative agent operational strategies: a skip-first strategy and a persistent search strategy. Several models including {\tt DS-Reasoner}, {\tt DS-v4-Flash}, and {\tt Qwen-3.5-Plus} explicitly choose to bypass challenging early stages and advance to subsequent tasks, justifying such decisions with time limits, iteration quotas, or perceived task difficulty. This strategy reduces resource consumption but sacrifices end-to-end pipeline completeness.
In contrast, other models adopt a persistent search strategy, repeatedly executing edit-compile-debug loops even when progress stalls. Although this persistence may occasionally resolve ambiguous implementation problems, it continuously expands the dialogue context, accelerates token consumption, and eventually leads to context exhaustion or budget overrun.

The divergence between the two strategies demonstrates that agent process quality depends not only on final task accuracy but also on adaptive effort allocation and long-horizon workflow scheduling under resource constraints. Moreover, such failure and behavioral patterns can only be exposed through \ramp{}’s serial dependency pipeline, which are largely invisible in conventional independent static benchmarks.

\subsection{Overall Utility}
Putting together mean reward, execution time, monetary cost, and token consumption, the Agent Efficiency Index (\aei{}) provides an operational measure of how efficiently an agent converts resources into useful engineering progress. This section therefore shifts from isolated outcome or resource measurements to a holistic view of utility, where a model is not only assessed by the reward it attains, but also by the time, budget, and context it spends to obtain that reward. \aei{} captures this productivity-oriented perspective and makes visible an important deployment trade-off between raw task utility and scalable resource discipline.
Higher \aei{} values indicate greater useful reward per unit of aggregate resource consumption, with time, cost, and token usage jointly treated as important resource dimensions.

Figure~\ref{fig:fig_radar} complements this scalar index with multi-dimensional profiles normalized efficiency profiles, where larger radii consistently indicate stronger stage progress, higher reward, shorter runtime, lower cost, or lower token consumption. These profiles reveal distinct efficiency tradeoffs across models that are obscured by single-value rankings.

\textbf{Finding 5: Agentic models exhibit sharply different rankings under \aei{} and mean reward.}
The highest mean reward is achieved by {\tt Opus-4.7} ($MR=93.39$), but its \aei is only 40.00 because it also attains the observed maximum on wall-clock time, API cost, and token usage (11{,}913.79\,s, \$126.24, and 218.86M tokens), yielding zero scores on all three inverted resource axes.
By contrast, {\tt GPT-5.5} achieves the highest AEI (81.57) with $MR=65.91$, combining the maximum pipeline stage ($S=3$) with mid-to-high scores on reward and all three resource dimensions.
{\tt DS-v4-Pro} reaches a higher mean reward (85.34) but a lower AEI (65.21), primarily because it consumes 6.17$\times$ more tokens than {\tt GPT-5.5} (136.76M vs.\ 22.18M).
{\tt Qwen3-Coder} illustrates the opposite pattern: strong time/cost/token scores at very low absolute cost (608.36\,s, \$0.05, 3.48M tokens), but limited stage progress and mean reward ($MR=25.97$, AEI 64.22).

This inversion is the central interpretive role of AEI.
It does not claim that a low-reward model is preferable in every setting.
Instead, it ranks models by balanced progress across mean reward and the three resource dimensions on a common $[0,100]$ scale.
For large-scale agent deployment, this distinction is essential: a model that obtains the best reward through orders-of-magnitude higher cost, latency, and context usage may be less attractive than a more modest model that delivers repeatable progress with far better resource discipline. For research settings focused on pushing capability boundaries or one-off high-stakes tasks, MR remains the more relevant metric.

\textbf{Finding 6: Top-tier LLM agents (e.g., Opus-4.7) bear excessive resource overhead for only marginal performance gains, with AEI clearly exposing such implicit inefficiency.}
The comparison between \texttt{GPT-5.5}, \texttt{DS-v4-Pro}, and \texttt{Opus-4.7} illustrates this effect.
\texttt{DS-v4-Pro} achieves a higher mean reward than \texttt{GPT-5.5} (85.34 vs. 65.91), but \texttt{GPT-5.5} has a higher AEI (81.57 vs. 65.21).
Their monetary costs are nearly identical (\$8.77 vs. \$8.68), so the efficiency gap is mainly driven by time and context usage: \texttt{DS-v4-Pro} uses 1.37$\times$ more time and 6.17$\times$ more tokens.
\texttt{Opus-4.7} pushes mean reward even higher, but its AEI falls to 40.00 because it uses 2.19$\times$ more time, 14.39$\times$ more money, and 9.87$\times$ more tokens than \texttt{GPT-5.5}.

Other models exhibit different forms of inefficiency.
\texttt{DS-v4-Flash} reaches a strong mean reward of 66.48, but its AEI is 56.40 because it consumes 157.85M tokens.
\texttt{Qwen-3.6-Max} reaches $R=56.72$ with AEI 66.75, primarily due to its high monetary cost of \$67.28.
These cases show that inefficient agent behavior is not one-dimensional.
Some agents are context-heavy, some are price-heavy, and some are slow.
AEI is useful because equal weighting across the five normalized dimensions prevents any one favorable reading from masking weakness on the others.

\textbf{Finding 7: Agentic models naturally fall into three distinct operational regimes in production workflows when measured by AEI.}
The AEI ranking reveals three distinct operating regimes for LLM agents in production workflows.
First, balanced productive agents exemplified by {\tt GPT-5.5} (AEI 81.57, $MR=65.91$) achieve the highest composite efficiency while maintaining solid task performance, making them the strongest choice for general-purpose production deployments under equal weighting assumption.
Second, marginally efficient solvers such as {\tt Qwen3-Coder} and {\tt DS-Chat} deliver mid-tier AEI values (64.22 and 60.36) at very low cost, making them suitable for early-stage prototyping and low-stakes tasks despite modest mean rewards.
Third, reward-heavy solvers including {\tt Opus-4.7}, {\tt DS-v4-Pro}, and {\tt DS-v4-Flash} obtain strong mean rewards but lower AEI (40.00, 65.21, and 56.40), making them valuable primarily when maximum task completion outweighs resource discipline.

Overall, AEI shows that current LLM agents differ not only in the reward they can obtain, but also in how efficiently they spend time, money, and context to obtain it. For long-horizon agent evaluation, this distinction is central: an agent that is powerful but expensive, slow, or context-hungry may be less productive in practice than a more modest agent that delivers reliable progress with disciplined resource use.
\section{Related Work}
Existing work on LLM benchmarking and evaluation can be broadly divided into categories of benchmarks, agent execution frameworks, and runtime infrastructure platforms.

\textbf{Repository-level coding benchmarks.}
Repository-level coding benchmarks have significantly advanced the evaluation of LLM-based software engineering systems. SWE-bench~\cite{jimenez2024swebench} established the paradigm of evaluating agents on real GitHub issues with execution-based patch verification, moving beyond isolated function synthesis benchmarks such as HumanEval~\cite{chen2021codexhumaneval}. SWE-bench Verified~\cite{chowdhury2024swebenchverified} further improved task reliability through human validation, while later extensions expanded the setting to multilingual repositories~\cite{zan2025swebenchmultilingual}, longer and more complex issue resolution~\cite{deng2025swebenchpro}, mobile development~\cite{yang2025swebenchmobile}, hardware engineering~\cite{hwbench2026}, and feature implementation~\cite{feabench2025}. Furthermore, recent advancements in 2026 have focused on refining the methodological rigor and scope of these benchmarks. For example, ATime-Consistent Benchmark~\cite{timeconsistent2026} introduces strict temporal boundaries to prevent data contamination from future repository states, while SWE-QA~\cite{sweqa2026} shifts the focus toward multi-hop codebase comprehension across dispersed files rather than patch generation alone. Similarly, USEbench~\cite{usebench2026} unifies various software engineering capabilities---ranging from code generation to program repair---into a comprehensive meta-benchmark. These benchmarks provide important execution-grounded signals, but their evaluation units remain largely independent. Each task typically starts from a clean repository state and ends with a single patch. As a result, they provide limited visibility into cumulative state evolution, intermediate artifact continuity, and cascading failure propagation. \ramp{} builds on the execution-grounded tradition of repository-level benchmarks, but shifts the evaluation target from isolated patch correctness to sustained progress along a serial dependency chain.

\textbf{Interactive agent benchmarks.}
Interactive benchmarks evaluate agents in environments that require tool use, multi-step reasoning, and interaction with external systems. AgentBench~\cite{liu2024agentbench} evaluates agents across diverse environments including operating systems, databases, and web tasks. GAIA~\cite{mialon2024gaia} focuses on general assistant competence under multi-step tool-augmented reasoning. OSWorld~\cite{xie2024osworld} and WebArena~\cite{zhou2024webarena} assess agents in realistic graphical and web-based environments, while Terminal-Bench~\cite{merrill2026terminalbench} provides execution-grounded command-line tasks. These benchmarks are valuable because they expose agents to broader interaction surfaces than static code-generation tasks. However, their strength is primarily breadth rather than serial depth. They generally do not require agents to maintain a continuously evolving software artifact across a strict multi-stage dependency pipeline. Consequently, they are less suited to measuring whether early-stage defects corrupt downstream execution, whether valid intermediate states can be preserved, or whether process costs accumulate in ways that affect deployment feasibility. \ramp{} fills this gap by making long-horizon dependency continuity and runtime observability central to the evaluation design.

\textbf{Paired and augmentation-oriented evaluation.}
Recent work has also studied evaluation through paired conditions. SkillsBench~\cite{li2026skillsbench} evaluates tasks under both vanilla and skill-augmented settings, showing that external skills do not uniformly improve agent performance. SWE-Skills-Bench~\cite{sweskillsbench2026} applies a similar methodology to software engineering tasks and finds that many public skills provide limited benefit. These works are methodologically relevant because they show the importance of comparing agent behavior under controlled intervention settings. However, their main objective is to measure augmentation efficacy. \ramp{} adopts a related paired-evaluation philosophy for a different purpose: failure decomposition. By comparing serial execution with and without resurrection, \ramp{} measures how much downstream diagnostic signal is hidden by upstream failure. This makes the intervention not an assistance mechanism for improving final scores, but a controlled assessment protocol for separating ``cannot reach'' from ``cannot solve'' in long-horizon workflows.

\textbf{Long-horizon and self-evolving agent evaluation.}
Long-horizon agent evaluation has become increasingly important as agents move from single-step coding assistance toward autonomous engineering workflows. Frontier-Eng~\cite{frontiereng2026} evaluates agents on realistic research and development tasks, while RE-Bench~\cite{rebench2024} and PaperBench~\cite{paperbench2025} study complex research-engineering settings. Voyager~\cite{wang2023voyager} demonstrates long-term skill accumulation in an open-ended environment. ProgramBench~\cite{programbench2026} further evaluates whether language models can reconstruct complete software systems from high-level specifications and behavioral signals, emphasizing holistic end-to-end software construction. In parallel, WildClawBench~\cite{wildclawbench2026} assesses agents in native-runtime CLI environments with realistic multi-step tool execution, highlighting the difficulty of maintaining persistent context in actual deployments. The extended interaction lengths in these environments also introduce novel vulnerabilities, as demonstrated by AgentLAB~\cite{agentlab2026}, which evaluates how agents succumb to compounding errors and long-horizon adversarial attacks during multi-turn trajectories. These works highlight the need to evaluate agents over extended trajectories, but they typically lack controlled intermediate-state restoration. When an agent fails early, the remaining trajectory is often either terminated or interpreted through the corrupted state produced by the failure. \ramp{} complements this line of work by introducing resurrection as a principled mechanism for continuing evaluation after upstream failure. This enables measurement of both strict end-to-end autonomy and latent downstream capability under corrected prerequisites, while also preserving process-level evidence about cost, context pressure, and recovery behavior.

Overall, existing benchmarks have made substantial progress in execution-grounded coding evaluation, interactive tool-use assessment, paired intervention studies, and long-horizon agent analysis. However, none of these lines of work simultaneously centers serial dependency, intermediate artifact continuity, resurrection-based failure decomposition, and deployment-oriented process metrics. \ramp{} is designed around this combination. Its goal is not only to determine whether an agent can solve a task, but to assess whether the agent can sustain useful engineering progress under realistic runtime constraints.

\section{Discussion on Limitations and Future Work}

Although \ramp{} provides a realistic and runtime-observable evaluation framework for long-horizon software development agents, several challenges remain insufficiently explored and warrant future investigation.

\textbf{Limitations}. First, \ramp{} currently focuses on compiler construction, which provides well-defined intermediate artifacts and deterministic grading, but may not fully represent other production domains such as machine learning engineering, DevOps automation, data pipelines, or web service maintenance. The methodology of serial dependency, resurrection, and process observation is transferable, but the empirical findings should not be assumed to generalize across all software engineering settings. Second, our experiments cover \cnt model tests, which may not be completely sufficient to reveal clear behavioral patterns, and thus can be limited for fine-grained statistical conclusions across model families, agent scaffolds, or repeated runs. Third, all experiments are performed using the OpenHands backend. This design improves internal consistency, but it also means that the observed behavior reflects the interaction between models and this specific agent framework. Different agent frameworks may produce different planning strategies, tool-use patterns, context-management behavior, or recovery dynamics. Fourth, \yatcc is publicly available, creating a potential contamination risk for frontier models that may have encountered parts of the codebase during training. Although \ramp{} primarily stresses serial continuity and artifact integration rather than memorization of isolated solutions, future private or temporally split workloads would provide stronger protection against this risk. Finally, the proposed AEI introduces a useful deployment-oriented view, but its equal weighting across reward, stage progress, time, cost, and tokens is necessarily subjective. Different deployment scenarios may prioritize latency, cost, reliability, or maximum reward differently, so AEI should be interpreted as one operational summary rather than a universal ranking criterion.

\textbf{Future Work}. These limitations point to several directions for future work. First, to broaden the evaluation scope by incorporating more models, more agent frameworks, and repeated runs under controlled settings, enabling clearer separation between model capability, scaffold design, and runtime orchestration effects. Second, to improve AEI so that it more objectively reflects agent process capability. This may involve learned or scenario-dependent weighting, Pareto-front analysis, robustness-aware scoring, or metrics that explicitly capture recovery behavior, context management, and productive tool use. Finally, future versions of \ramp{} should include harder workload variants with reduced or removed \yatcc code scaffolding. The current scaffolded setting evaluates whether agents can complete and evolve an existing compiler pipeline; removing scaffolding would further test architectural reasoning, from-scratch implementation ability, and the capacity to maintain system-level consistency with less structural guidance. Together, these extensions would make \ramp{} a stronger foundation for studying not only whether agentic models can solve production-like tasks, but also whether they can do so robustly, efficiently, and with deployable engineering discipline.

\section{Conclusion}
Static and isolated benchmarks are increasingly insufficient for faithfully evaluating agentic model capabilities. Real-world software development workflows are long-horizon, stateful, dependency-driven, and resource-constrained, making final task accuracy alone inadequate for assessing practical deployability.
We therefore present \ramp{}, a production-grounded assessment infrastructure built on \yatcc{} compiler-construction tasks. Through serial workflow execution, resurrection-based failure decomposition, and multi-dimensional runtime metrics, \ramp{} assesses not only task outcomes, but also execution behavior, artifact continuity, resource efficiency, and downstream recoverability.
Our findings reveal substantial gaps between benchmark performance and production robustness, demonstrating that practical utility cannot be judged by reward alone.
A meaningful assessment must look into the agent's process: how it spends time, cost, and context; how it handles failures; and whether it can maintain progress across dependent engineering stages.

\section*{Acknowledgment}
We would like to sincerely thank all contributors, collaborators, students, and users who participated in the design, development, deployment, operation, and continuous evolution of the \yatcc platform and its ecosystem. \yatcc is the result of extensive collective efforts spanning infrastructure engineering, platform operations, educational practice, and community-driven innovation. Their dedication, technical insights, feedback, experimentation, and long-term support have been instrumental in transforming \yatcc from an initial course teaching tool into a practical AI-native serving platform.

\bibliographystyle{ACM-Reference-Format}
\bibliography{references}

\end{document}